\documentclass[aps,pra,twocolumn,superscriptaddress]{
revtex4-2}
\usepackage{amsmath}
\usepackage{amssymb}
\usepackage{graphicx}
\usepackage{hyperref}% add hypertext capabilities
\usepackage{color}
\usepackage{bm}
\usepackage{braket}
\usepackage{epstopdf}
\usepackage{amstext}
\usepackage{amsthm}
\usepackage{amsfonts}
\usepackage{latexsym}
\usepackage{array}
\usepackage{xfrac}
\usepackage{lipsum}
\usepackage[toc,page]{appendix}
\usepackage{subfigure}

\usepackage[normalem]{ulem}

\renewcommand{\vec}[1]{{\boldsymbol{#1}}} % for vectors

\newcommand{\dbar}{{\mathchar'26\mkern-11mud}}
\begin{document}
	
	%\preprint{This line only printed with preprint option}
	\title{Squeezing Bose-Bose liquid drops}
	\author{A. Sanuy}
	\affiliation{Departament de F\'{\i}sica, 
		Universitat Polit\`ecnica de Catalunya, 
		Campus Nord B4-B5, E-08034 Barcelona, Spain}
	\author{R. Bara\v{c}}
	\affiliation{Faculty of Science, University of Split, Ru$\,\dbar$era 
		Bo\v{s}kovi\'{c}a 33, HR-21000 Split, Croatia}
	\author{P. Stipanovi\'c}
	\affiliation{Faculty of Science, University of Split, Ru$\,\dbar$era
		Bo\v{s}kovi\'{c}a 33, HR-21000 Split, Croatia}
	\author{L. Vranje\v{s} Marki\'c}
	\affiliation{Faculty of Science, University of Split, Ru$\,\dbar$era
		Bo\v{s}kovi\'{c}a 33, HR-21000 Split, Croatia}
	\author{J. Boronat}
	\affiliation{Departament de F\'{\i}sica, 
		Universitat Polit\`ecnica de Catalunya, 
		Campus Nord B4-B5, E-08034 Barcelona, Spain}
	\date{\today}% It is always \today, today,
	%  but any date may be explicitly specified
	
	\begin{abstract}
	 We explore ultradilute Bose-Bose liquid droplets squeezed by an external 
harmonic potential in one spatial direction. Our theoretical study is 
based on a functional that is built using quantum Monte 
Carlo results of the bulk phase and incorporates finite-range effects. 
A characteristic feature of these drops is the existence of a critical atom 
number, that is the minimum number of particles to have a many-body bound 
state. We report results on the critical atom numbers for different magnetic 
fields and applying  confinement strengths towards a quasi-two-dimensional 
setup. In the 
regime where the local density approximation is expected to be valid, we find 
that the critical atom number decreases 
linearly with the harmonic oscillator length of the confining 
potential. With the largest squeezing explored in our work, we predict stable 
drops at the level of one thousand atoms. 
Our functional reduces the critical numbers for any confinement and 
applied magnetic field with respect to the estimations based on the 
Lee-Huang-Yang model.  We observe saturated drops when the number of atoms in 
the drop is much larger than the critical value, their central density being 
higher for the quantum Monte Carlo functional than for the Lee-Huang-Yang 
one. 
%When the number of particles is close to the critical number 
%increasing the squeezing makes the droplet less flat. On the other hand, when 
%the droplets reach saturation density, similarly to the helium droplets, 
%squeezing results in flatter droplets.
	\end{abstract}
	
	% PACS, the Physics and Astronomy
	% Classification Scheme.
	%\keywords{Suggested keywords}%Use showkeys class option if keyword
	%display desired
	\maketitle
	
	\section{\label{intro} Introduction}

Ultradilute liquid droplets were  
 predicted by Petrov~\cite{petrov2015quantum} and soon after confirmed in 
experiments with a Bose-Bose mixture of two hyperfine states of 
$^{39}$K~\cite{cabrera2018quantum,semeghini2018self}. These many-body bound 
states arise  from the interplay between attractive interspecies 
and repulsive intraspecies interactions and, in this sense, they differ 
from the unicomponent classical or 
quantum Helium droplets that result from the interplay between the 
repulsive short-range and attractive long-range components of the interatomic 
potential. 
The delicate balance between attractive and repulsive interactions 
produces Bose-Bose drops which are extremely dilute, with densities that are 
orders of magnitude smaller that the ones of liquid Helium.  
In fact, very dilute self-bound systems with dipolar atoms were experimentally noticed even before,
arising from the interplay of short-range repulsion and
attractive head-to tail dipolar moments~\cite{Ferrier-Barbut,Schmitt2016,Kadau2016}. Further
experimental efforts have resulted in heterogeneous 
Bose-Bose droplets~\cite{derrico2019observation,Guo21} of $^{41}$K-$^{87}$Rb  and
$^{23}$Na-$^{87}$Rb. The emergence of these new ultradilute liquids has been recently
reviewed in Refs.~\cite{Luo2020,Bottcher2021}.

According to mean-field theory, binary mixtures of Bose-Einstein condensates 
become unstable against collapse when the attractive interspecies interaction 
overcomes the repulsive contact potential between atoms~\cite{pethick2008bose}. 
However, in the ultradilute liquid phase the mean-field collapse is avoided if 
beyond-mean-field first-order perturbative corrections, in the form of the 
Lee-Huang-Yang (LHY) energy functional~\cite{lee1957eigenvalues,LARSEN196389}, 
are included. This correction, which takes into account quantum fluctuations, is 
repulsive and thus it can stabilize attractive Bose-Bose 
mixtures~\cite{petrov2015quantum}. Going beyond the LHY corrections, the 
formation of Bose-Bose droplets was also observed using the 
first-principles diffusion Monte Carlo (DMC) 
method~\cite{Cikojevic18ultradilute}.

 In the Bose-Bose mixture, the LHY functional suffers from an intrinsic 
inconsistency, with the appearance of an imaginary term in the energy for 
densities different from the equilibrium one. It has been suggested that this 
drawback can be eliminated by introducing bosonic 
pairing~\cite{Hu2020pairing,Hu2020PRA} or time-dependent Hartree-Fock-Bogoliubov 
theory~\cite{Boudjema21}. This is at variance with a single-component Bose gas, 
where no such imaginary term appears and where the LHY term is able to explain 
accurately the departure of the energy from the mean-field 
prediction~\cite{Giorgini99,Rossi2014}. Furthermore, observed 
properties~\cite{cabrera2018quantum} of an ultradilute mixture of two hyperfine 
states of $^{39}$K  showed deviations from the predictions of the theory based 
on interactions described solely in terms of $s$-wave scattering 
lengths~\cite{petrov2015quantum}, suggesting the inclusion of finite-range 
corrections beyond LHY. The dependence of the energy on the effective range was 
noticed in a variational hypernetted-chain Euler-Lagrange 
calculation~\cite{Staudinger}, and confirmed later on by the study of  
ultradilute liquids with the DMC method~\cite{cikojevic2019universality}. 
Interestingly, those DMC simulations  showed that, by including the effective 
range, the universality of the theory can be 
extended~\cite{cikojevic2019universality,cikojevic2020finite}. The improved 
energy functionals with built-in knowledge of both $s$-wave scattering lengths 
the effective ranges, allowed   approaching  experimental results for the 
critical atom number~\cite{cikojevic2020finite}, as well as addressing 
excitation spectra~\cite{cikojevic2020excitation,Hu2020excitation} and the study 
of the previously observed~\cite{Ferioli19}   dynamics of equilibration and 
collisions of  $^{39}$K droplets~\cite{cikojevic2021collision}.

The interest in ultradilute droplets was very soon extended to low-dimensional
systems, where it was found that in two dimensions (2D) the liquid phase is 
formed whenever the intraspecies interactions are repulsive and the interspecies 
one is attractive~\cite{petrov2016ultradilute}. LHY correction was also studied 
in the crossover from three to two  and from three to one 
dimensions~\cite{Ilg,Zin18}, which provided conditions for the use of 
three-dimensional (3D) local-density functionals to study droplets in 
confinement.  Recently, Bose-Bose droplets squeezed by an external harmonic 
potential in one spatial direction have been studied using the modified gapless 
Hartree-Fock-Bogoliubov method~\cite{Zin_2022}, without assuming 
local-density approximation in the direction of squeezing~\cite{Zin_2022_PRA}.

Previous research shows that departures from mean field + LHY predictions 
(MFLHY) 
are stronger when the mixtures are compressed~\cite{cikojevic2020finite}. It 
would be particularly interesting, for the range of the experimentally 
accessible magnetic fields, to investigate how the critical atom numbers change 
as the droplet is squeezed, as well as how the density profiles of drops change. 
The properties of more strongly compressed Bose-Bose mixtures, where quantum 
fluctuations will be even more prominent, have not been studied using density 
functional theory incorporating information on the effective range. In the 
present work, we extend our previous study of confined 
drops~\cite{cikojevic2020finite} to regimes of stronger confinements, reaching 
the boundary of the expected validity of the local-density 
approximation~\cite{Ilg}.

The paper is organized as follows. In Section II,  we briefly introduce the 
density functional approach used in our study, as well as the MFLHY functional. 
The results on the droplet energies, critical atom numbers and droplet profiles, 
for different magnetic fields and squeezing strengths are presented in Section 
III. Finally, Section IV is devoted to summary and conclusions.
	
	\section{\label{methods} Methods}

We study  Bose-Bose mixtures  of two hyperfine states of $^{39}$K  at 
zero temperature, using the density functional approach. The many-body wave 
function is built as a product of single-particle  orbitals $\psi_{1,2}$,
\begin{equation}
	\label{eq:dft_wf}
	\Psi(\vec{r}_1, \vec{r}_2, \ldots, \vec{r}_N)= 
	\prod_{i=1}^{N_1} \psi_1(\vec{r}_i)\prod_{j=N_1+1}^{N_2} \psi_2(\vec{r}_j),
\end{equation}	
where $N_1$ ($N_2$) are the number of particles of the first (second) component. 
The single-particle wave functions are 
obtained solving the Schr\"odinger-like equation of motion for each species,
\begin{equation}
	\label{eq:timegp}
	i\hbar\dfrac{\partial \psi_i}{\partial t} = \left(-\dfrac{\hbar^2}{2m} 
	\nabla^2 + V_{\rm ext}(\vec{r}) + \dfrac{\partial \mathcal{E}_{\rm 
			int}}{\partial \rho_i}\right) \psi_i \ ,
\end{equation}
where $V_{\rm ext}$ is an external potential acting on the system, $m$ is the  
$^{39}$K atomic mass, and
$\mathcal{E}_{\rm int}$ is an energy per volume term that accounts for the 
interparticle correlations.	The density of each component is $\rho_i=
|\psi_i|^2$.
In the following, we consider  $^{39}$K mixtures at   
the optimal relative atom concentration yielded by mean-field theory, namely 
$N_1 / N_2 = \sqrt{a_{22} / a_{11}}$~\cite{petrov2015quantum}. It is worth 
noticing that this optimal ratio between numbers of particles of both type has 
been confirmed in DMC simulations~\cite{cikojevic2020finite}.
Keeping this ratio fixed, the coupled differential equations (\ref{eq:timegp}) 
can be reduced to a single one, which is a function of the total density 
$\rho$. 

For these mixtures, it was shown that the energy per atom, using the 
data obtained from the DMC calculations, can be  accurately written 
as~\cite{cikojevic2020finite}       
\begin{equation}
	\dfrac{E}{N} = \alpha \rho + \beta \rho^{\gamma} \ .
	\label{eos}
\end{equation}
The parameters $\alpha$, $\beta$, and $\gamma$ 
were determined by fits to the diffusion Monte Carlo results obtained using  
the model potentials satisfying both the $s$-wave scattering lengths and 
effective ranges. The values of these parameters for the range of experimentally 
relevant 
magnetic fields, from $B=56.23$ G to $B=56.639$ G, are reported  in 
Ref.~\cite{cikojevic2021abinitio}. This allows us to obtain the part of the 
functional describing interparticle correlations through the relation 
$\mathcal{E}_{\rm{int}}=\rho\,E/N$.

For a couple of reference cases, we performed calculations using the 
MFLHY functional. Working at the optimal concentration, 
the energy per particle can be written as in Eq. (\ref{eos}), 
\begin{equation}
	\label{eq:mflhyeos}
	\dfrac{E/N}{|E_0|/ N} = -3\left(\dfrac{\rho}{\rho_0}\right) + 2 
	\left(\dfrac{\rho}{\rho_0}\right)^{3/2} \ ,
\end{equation}
with $E_0/N$ the energy per atom at equilibrium, 
\begin{equation}
	\label{eq:en0}
	E_0 / N = -\dfrac{25 \pi^2 \hbar^2 |a_{12} + \sqrt{a_{11} 
			a_{22}}|^3}{768m a_{22} a_{11} \left(\sqrt{a_{11}} + \sqrt{a_{22}}\right)^6} 
\end{equation} 
and $\rho_0$ the equilibrium density,
\begin{equation}
	\label{eq:rho0}
	\rho_0  = \dfrac{25 \pi }{1024 a_{11}^3} \dfrac{\left(a_{12}/a_{11} + 
		\sqrt{a_{22}/a_{11}}\right)^2}{\left(a_{22}/a_{11}\right)^{3/2}\left(1+\sqrt{a_{
				22}/a_{11}}\right)^4} \ .
\end{equation}
In Eqs. (\ref{eq:en0}) and (\ref{eq:rho0}), $a_{ij}$ are the three different 
$s$-wave scattering lengths, whose values for the selected magnetic fields are 
given in Ref.~\cite{cikojevic2020finite}. 

For the external potential, we have chosen a harmonic confinement in the $z$ 
direction, $V_{\rm ext}=m\omega^2_zz^2/2$, with $\omega_z=\hbar/(m 
a_{\rm ho}^2)$. In the reference  experiment~\cite{cabrera2018quantum},  the same confinement potential was applied with $a_{\rm ho}$=0.639$\,\mu$m. As we 
are interested in the effect of squeezing on the properties of the drops,  
we have taken further compression, $a_{\rm ho}$=$f\times$0.639$\mu$m with 
$f=$0.9, 0.75, 0.5, and 0.25.	

We are interested on finding the minimum energy configuration. Therefore, we 
solve  the differential equation 
(\ref{eq:timegp}) in imaginary time $\tau \equiv i t$. We propagate  
the wave function $\psi$ 
with the time-evolution operator
\begin{equation}
	\psi(\tau + \Delta \tau) = e^{- \Delta \tau \, H } \psi(\tau) \ .
\end{equation}
To this end, we have implemented a three-dimensional numerical solver based 
on the Trotter decomposition of the time evolution 
operator~\cite{chin2009any} with second-order accuracy  
in the time step $\Delta \tau$, as follows 
\begin{equation}
	e^{- \Delta \tau \, H}  \approx e^{-\Delta \tau V(\vec{R}') / 2}  
e^{-\Delta \tau K } e^{-\Delta \tau V(\vec{R})/ 2} + \mathcal{O}((\Delta 
\tau)^2) \ ,
\end{equation}
with $K$ and $V$ the kinetic and potential terms in 
Eq. (\ref{eq:timegp}).
	
	\section{\label{results} Results}
	
\begin{figure}
	\centering
	\includegraphics[width=\linewidth]{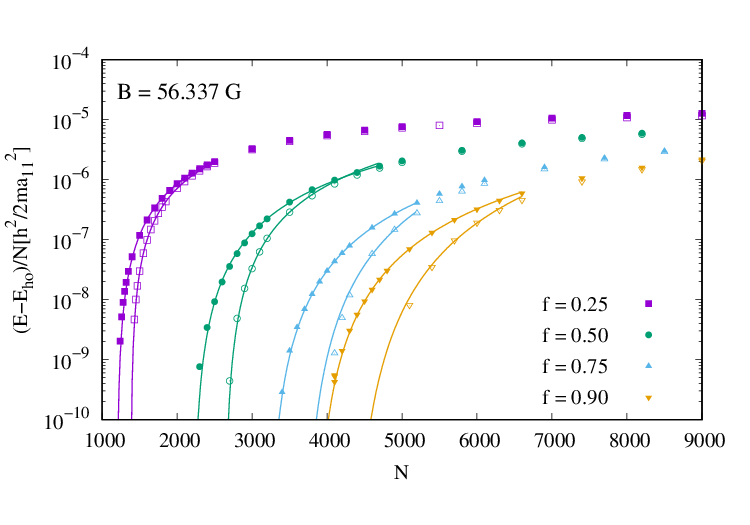}
	\caption{Total energy of the droplet per particle as a function of the 
number of particles for the magnetic field $B=56.337$ G and several 
confinements, described by harmonic oscillator lengths  $a_{\rm ho}=f \times 
0.639$ $\mu$m. The energies are obtained using both the QMC (full symbols) and 
MFLHY (empty symbols) energy functionals.   }
	\label{fig:Ncrit_337}
\end{figure}	

As commented in the Introduction, one of the characteristics of Bose-Bose drops 
is the requirement of the number of particles being larger than a certain 
critical atom number $N_c$. This number is very sensitive to the applied 
magnetic field and to the transversal confinement and, moreover, to the theory 
applied to obtain it. We calculate the energy of the drops and estimate the 
minimum atom number that keeps the drop self bound, that is the number that 
gives zero binding energy. This study is carried out 
for each confinement strength, described by the harmonic oscillator length 
$a_{\rm ho}= f\times 0.639$ $\mu$m. After subtracting the single 
particle energy contribution due to the harmonic oscillator potential, the 
energies per particle of drops, as a function of the atom number, are fitted to 
the functional form 
\begin{equation}
\epsilon(N)=10^{-\alpha / (N-N_c)^\beta}    \ , 
\label{eq:fit}
\end{equation}
from which the critical atom number $N_c$ is determined. 
The dependence of the energies on the cluster size and the fitting model 
(\ref{eq:fit}) are presented in Fig.~\ref{fig:Ncrit_337} for the magnetic field 
strength $B=56.337$ G, using the two density functionals, QMC  and MFLHY. For 
both approaches, squeezing increases the binding strength and reduces 
$N_c$. The MFLHY density functional results with larger 
values for $N_c$ with respect to the QMC functional, in the range from 13\% to 
26\%. The functional form used for fitting describes well the energies for  
numbers of particles close to $N_c$, but the particular value of $N_c$ depends 
only slightly on the chosen range of clusters included in the fit procedure. 
The reported error bars for $N_c$ in Table~\ref{tbl:Ncrit} includes the spread 
of $N_c$ values obtained by statistically compatible  fits corresponding to 
different ranges of cluster sizes.

\begin{table*}
	\caption{Critical number of atoms needed to form a bound droplet for 
different magnetic fields and confinement strengths, described by the harmonic 
oscillator length $a_{\rm ho}=f\times 0.639$ $\mu$m. The first line are obtained using the QMC
functional, while the data in second line are obtained by the MFLHY functional. 
The data for $f=1$ are from the Ref.~\cite{cikojevic2020finite}. The third row
for $f=1$ are experimental data from Ref.~\cite{cabrera2018quantum}.}
	\label{tbl:Ncrit}
	\begin{tabular}{lccccccc}
		\hline
		& &  &  & $B($G$)$ & &\\
		\cline{2-8}
		$f$~~~~   & ~56.230~  & ~56.337~ & ~56.395~ & ~56.453~   &   ~56.511~  &   ~56.547~ &    ~56.639~\\
		\hline
		1  & 3500 & 4200 & 5000  & 6000 &   7000 &   8500 &    11300\\
		   & 4650 & 5570 & 6200 & 7000  &   8050  &  9800 &  12700 \\
		   & -    & 3420(855) & 3421(855) & 4373(1093) & 7052(1763) & 9217(2304) & 13819(3455) \\
		\hline
		0.9 &  3100(50)    &      3800(50) &  4650(100) & 5400(100) & 6250(50)   &   7400(50) & 10300(100)     \\
		    &      &  4300(100) &   &            &         & 8400(100)         &       \\
		\hline
		0.75 &  2600(50)  &   3200(50) &   3900(50)  &   4550(50) & 5200(100)   &    6100(50)   & 8350(100)         \\
		     &     &  3600(50) &       &            &     &        &           \\
		\hline
		0.5 &   ~1750(50)~ &  ~2200(50)~ & ~2550(50)~ & ~3000(50)~ & ~3400(50)~ & ~4200(50)~ & ~5700(100)~ \\
		   &               &  2600(50) &            &            &             &            &           \\
		\hline
		0.25 &  970(30) &   1200(30) & 1450(30) & 1700(30) & 2050(50) & 2550(50) & 3850(50) \\
		     &          &   1400(50) &          &          &          &          &          \\
		\hline
	\end{tabular}
\end{table*}

The results for critical numbers of particles obtained with the QMC functional 
are presented in Table~\ref{tbl:Ncrit} and compared with obtained MFLHY 
results for selected cases.  For completeness, the results from 
Ref.~\cite{cikojevic2020finite} for $f=1$ are added as well. To facilitate 
interpretation, we note that lower values of magnetic field correspond to larger 
values of $|\delta a|$, with $\delta a=a_{12}+\sqrt{a_{11}a_{22}}$. 
For instance, droplets in a $B$=56.230 G magnetic field, with $\delta 
a=-3.45$ \AA\, are self bound with atom numbers smaller than in  
$B$=56.639 G, with $\delta a=-1.29$ \AA\,. As a general rule, the binding 
energy increases with  $|\delta a|$ and so, with decreasing $B$.

	\begin{figure}
	\centering
	\includegraphics[width=\linewidth]{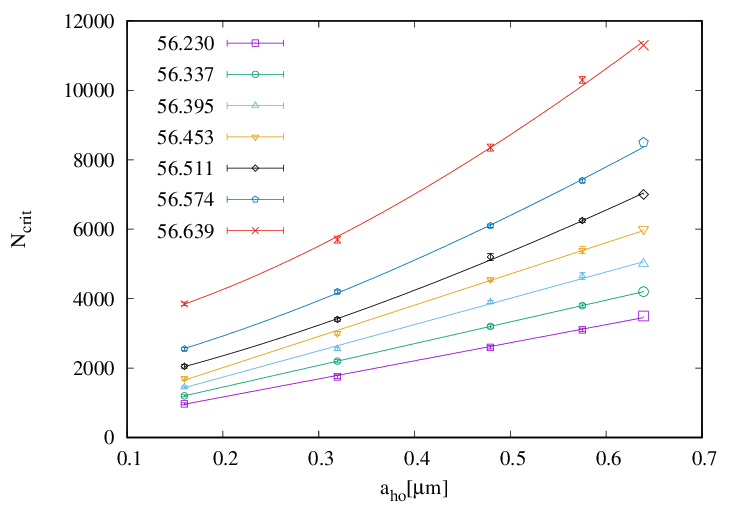}
	\caption{Critical atom number $N_{c}$ needed to form a 
bound droplet as a function of the harmonic oscillator wavelength $a_{\rm ho}$, 
for different magnetic fields (in G). The error bars are in most cases smaller 
than the symbol size. The results from Ref.~\cite{cikojevic2020finite}, 
corresponding to the largest $a_{\rm ho}$, are marked with larger symbols. The 
lines represent least-squares fits to data.}
	\label{fig:Ncrit1}
\end{figure}

In Fig.~\ref{fig:Ncrit1}, we show the dependence of the critical atom number on 
the oscillator length, for the range of magnetic fields included in 
Table~\ref{tbl:Ncrit}. As one can see, $N_c$ increases with $a_{\rm ho}$ 
following an approximately linear behavior.
Indeed, the linear function fits well the results for magnetic fields in 
the range $B$=56.230--56.453 G, while for larger values of $B$  
better results are obtained with the functional form $N_{c}=N_0+c a_{\rm 
ho}^d$, with $d$ departing from 1 as the magnetic field is increased. In all 
cases, $N_c$ for the lowest studied $a_{\rm ho}$ are $\sim$70\% lower 
than the values for $f=1$ (the largest $a_{ho}$ in 
Fig. \ref{fig:Ncrit1}), which were reported  in 
Ref.~\cite{cikojevic2020finite}. 
%This approximate linear behavior can be a 
%useful guide to experimentalists trying to obtain smaller cluster sizes.  
We note that in the range of squeezing 
studied  $a_{11} \ll a_{\rm ho}$, which is a prerequisite for using a 3D 
approximation. Two studies of the crossover from three to two dimensions in a 
system with periodic boundary conditions determined an additional bound for the 
validity of local-density approximation~\cite{Ilg,Zin18}. Particularly, in the 
case of harmonic confinement it was found that the 3D functional can be used 
when $\kappa=n_{2D}a_sa_{\rm ho}\gtrsim 1$, where $n_{2D}$ is the density 
integrated over the squeezing direction and $a_s$ is the scattering 
length~\cite{Ilg}. Using the value of the central density of the large clusters, 
which corresponds to the saturation density of the bulk, and taking the largest 
scattering length $a_{11}$, we have estimated the $\kappa$ parameter and 
obtained that, in most cases, $\kappa \gtrsim 1$.  The exceptions are the 
strongest confinements for the smaller $|\delta a|$ (larger magnetic fields), 
where $\kappa$ is 0.2, 0.3, 0.4 and 0.6 for $f=0.25$, going from $B$=56.639 to 
$B=56.453$, and $\kappa=0.6$ for $f=0.5$ and $B=$ 56.639, while for other 
parameters it ranges from 1 to almost 22. It was found that the correction to 
the local density approximation for systems approaching 2D lowers the 
ground-state energy~\cite{Ilg}. Therefore, it is possible that the density 
functional for $\kappa \lesssim 1$ is missing a confinement-induced 
correction, which could explain the departure from the linear 
behavior observed in Fig.~\ref{fig:Ncrit1}.   Also, as expected using a 3D 
functional, the lines do not extrapolate to the correct 2D limit, $N_0$=2, in which clusters of all sizes are bound, as 
reported in Ref.~\cite{petrov2016ultradilute}. 

\begin{figure}
	\centering
	\includegraphics[width=\linewidth]{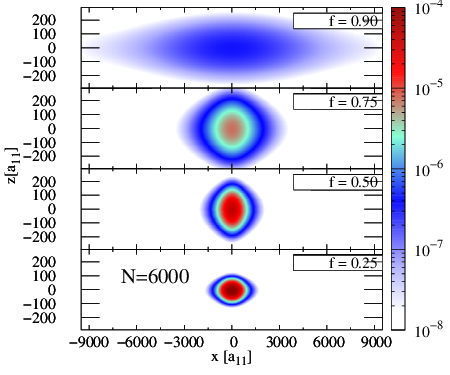}
	\caption{2D density profiles $\rho(x,0,z)$ of drops with 6000 atoms 
	for the magnetic field $B=56.453$ G and different squeezing strengths, 
described by the harmonic oscillator length of $a_{\rm ho}=f\times 0.639$ 
$\mu$m. The profiles are obtained using the QMC functional. Notice the 
different scales in the direction of the squeezing (vertical) and perpendicular 
to it. }
	\label{fig:rho_56_453_2d}
\end{figure}

The density profiles of the drops are relevant quantities to know their shape 
and size. By changing the interactions and the squeezing, one can study the 
effects of 
confinement and interactions on the size of the clusters. 
Figure~\ref{fig:rho_56_453_2d} shows the density plots $\rho(x,0,z)$ of a drop 
with 6000 atoms for several squeezing strengths. As the squeezing increases, 
the half-width of the density profile in the direction of the squeezing, 
$\Delta_{\parallel}$, goes from 257$a_{11}$ for $f=0.9$ to 71$a_{11}$ for 
$f=0.25$. Therefore, even for the tightest confinement the density profile width 
is significantly larger than the scattering lengths.  We observe that the drops 
are much more extended in the direction perpendicular to squeezing. The chosen 
drop with $N=6000$ particles, in the case of $f=0.9$ is close to the threshold 
of binding (estimated  $N_c=$5400(100)) and spreads to several thousands of 
$a_{11}$ in the direction perpendicular to the confinement (half-width 
$\Delta_{\perp}=$6600 $a_{11}$, while the width at 10\% of central 
density is 15000$a_{11}$). For the strongest squeezing, the perpendicular size is 
reduced by an order of magnitude ($\Delta_{\perp}$=680$a_{11}$, while the width 
at 10\% of central density is 1370$a_{11}$). The squeezing for the studied range 
of parameters thus makes the small weakly bound drops less flat, changing the 
aspect ratio $\Delta_{\perp}/\Delta_{\parallel}$ from 25.7 for the weakest 
binding to 9.6 for the strongest one.

\begin{figure}
	\centering
	\includegraphics[width=\linewidth]{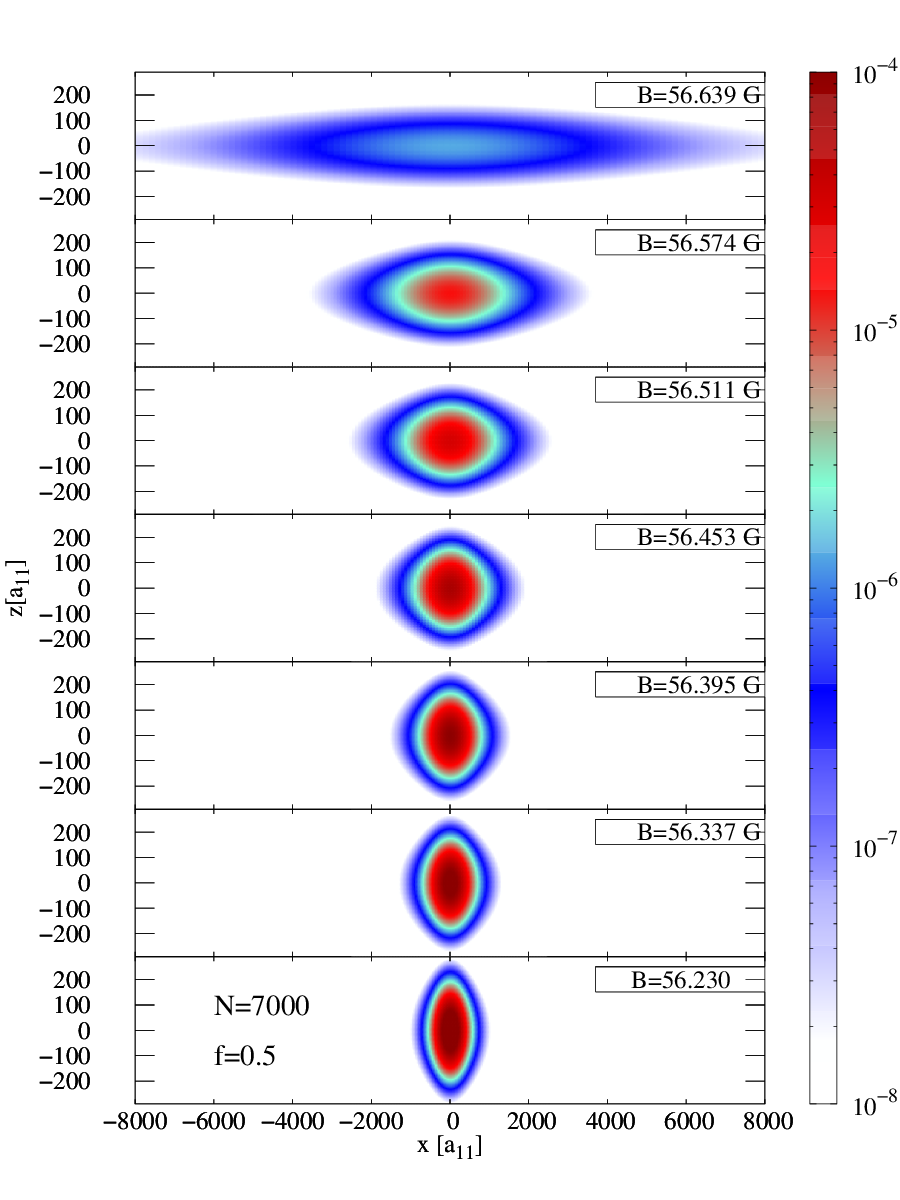}
	\caption{2D density profiles $\rho(x,0,z)$ of drops with 7000 atoms for 
	different magnetic fields with a fixed squeezing $f=0.5$. 
Results are obtained using the QMC functional. Notice the different scales in 
the direction of the squeezing (vertical) and perpendicular to it.  } 
	\label{fig:rho_0_5_B_QMC_2d}
\end{figure}

It is interesting to explore magnetic field  effects on the cluster size for a 
particular number of particles. 
Figure~\ref{fig:rho_0_5_B_QMC_2d}  shows the density profiles of drops 
$\rho(x,0,z)$ 
with 7000 atoms for different strengths of the magnetic field, and a fixed  
squeezing $f=0.5$. With the decrease of magnetic field, the resulting 
attractive interaction ($|\delta a|$) increases and thus the drops become more 
strongly bound and less flattened. Namely, as the attraction is increased, the 
spread of $\rho(x,0,0)$ distributions is significantly decreased, while the width 
of $\rho(0,0,z)$ distributions is even slightly increased. In the case of the 
weakest binding ($B$=56.639 G), the half-width of the density profile in the 
direction of squeezing is $\Delta_{\parallel}=132a_{11}$, while the 
perpendicular one is $\Delta_{\perp}=5000a_{11}$, resulting in 
$\Delta_{\perp}/\Delta_{\parallel}\approx 38$. For the strongest binding 
($B$=56.230 G) the central density increases 166 times and the size is reduced 
so that  $\Delta_{\parallel}=150a_{11}$, $\Delta_{\perp}=400a_{11}$, and 
$\Delta_{\perp}/\Delta_{\parallel}\approx 2.7$.

\begin{figure}
	\centering
	\includegraphics[width=\linewidth]{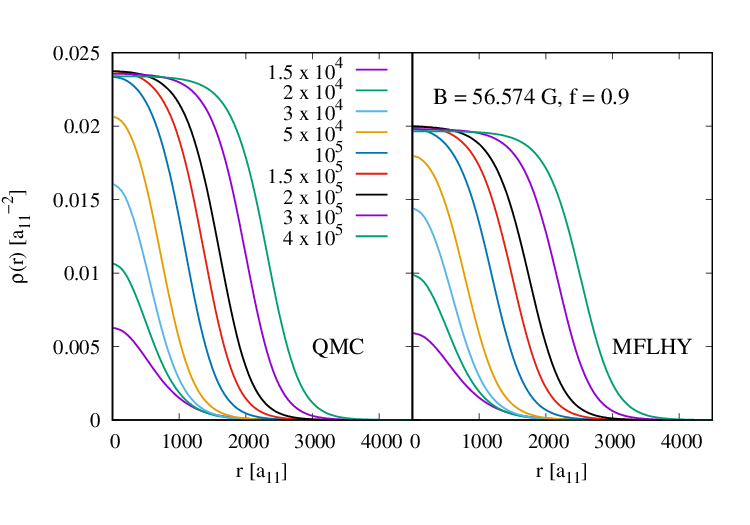}
	\caption{The density profiles of the drops in the direction perpendicular 
	to squeezing,  $r=\sqrt{y^2+z^2}$, obtained with the QMC  
(left panel) and MFLHY functional (right panel). The squeezing strength is 
$f=0.9$.   }
	\label{fig:rho_xy_0_9_574_QMC}
\end{figure}

\begin{figure}
	\centering
	\includegraphics[width=\linewidth]{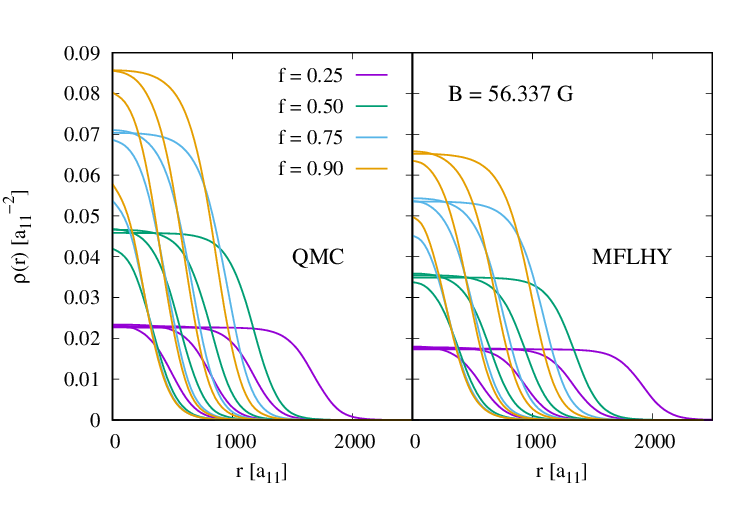}
	\caption{The density profiles of the drops in the direction perpendicular 
	to squeezing,  $r=\sqrt{y^2+z^2}$, calculated by the QMC (left) 
and MFLHY (right) functional, for $B=56.337$ G.  For all $f$ squeezing 
intensities,  four droplet sizes are shown going from left to right, $2\times 
10^4, 5\times 10^4, 10^5$ and $2\times 10^5$.  },
	\label{fig:rho_xy_56_337_QMC_LHY}
\end{figure}

Finally, we present the evolution of the density profiles with  
the number 
of particles $N$ in Figs.~\ref{fig:rho_xy_0_9_574_QMC} and 
\ref{fig:rho_xy_56_337_QMC_LHY}. We integrate over $z$ (the direction of the 
squeezing) and plot the slice in the perpendicular plane which corresponds to 
the distribution $\rho(r)$, with $r=\sqrt{x^2+y^2}$. The profiles are 
normalized in such a way that $2\pi \int \rho(r) rdr=N$.
In both figures, as the number of particles is increased we observe 
the saturation of the central density, a characteristic property of liquid 
drops. It is worth noticing the large number of particles in the drop to 
reach saturation, a consequence of the small binding energy of these so dilute 
drops.
For $B=56.574$ G, shown in Fig.~\ref{fig:rho_xy_0_9_574_QMC},  
the saturation density predicted by the MFLHY functional is about 15\% smaller 
than the QMC one. On the other hand, and similarly to the case of the Helium 
droplets~\cite{Stringari87,he2d}, for some intermediate size the central density 
is slightly larger than the saturation density of the equilibrium bulk, a 
consequence of the surface tension of drops. 

The saturation density increases with the interparticle attractive interaction 
strength, but 
qualitatively the density profiles evolve in the same way with the increase of 
the droplet size, as shown in Fig. \ref{fig:rho_xy_56_337_QMC_LHY} for the case 
of $B=56.337$ G. The plots show four droplet sizes for particle numbers  
$2\times 10^4$, $5\times 10^4$, $10^5$, and $2\times 10^5$ (from left to 
right) and for four different confinement strengths. With the increase of the 
squeezing, the saturation density decreases and it is reached for smaller 
number of particles. 
Unlike the case of droplets with small number of particles shown in Figs. 
\ref{fig:rho_56_453_2d} and
\ref{fig:rho_0_5_B_QMC_2d}, as the squeezing is 
increased, the spread of droplets shown in Fig. 
\ref{fig:rho_xy_56_337_QMC_LHY}, 
in the direction perpendicular to squeezing, increases. The difference between 
the QMC and MFLHY functional predictions grows as the interaction strength is 
increased, reaching around 24\% difference in central density, accompanied with 
the corresponding change in the spread of the cluster. 

\section{\label{discussion} Summary and Conclusions}

We have studied the effects of increased confinement on the binding properties, size, and shape
of weakly bound Bose-Bose droplets, using a QMC-based density functional.
By squeezing the droplets beyond the strength used in the experiment
\cite{cabrera2018quantum}, we have observed the decrease of the critical atom number
needed to form a droplet, reaching less than 1000 particles for the
case of the strongest interparticle interaction ($B=56.230$ G) and tightest
studied confinement, $a_{\rm ho}=0.25\times 0.639$ $\mu$m. Our results show that, in the
observed range of squeezing, the lowering of the critical atom number is
approximately linear with the harmonic oscillator length. Deviations from
the linear behavior appear for the larger magnetic fields (smaller $|\delta 
a|$) 
and the strongest squeezing, signaling the approach of the 3D-2D crossover, where
the local density approximation used may require corrections. It would be useful 
to perform the full Quantum Monte Carlo calculation in these cases and asses the 
validity of the functional.

 Clusters with the number of particles of the order of $N_c$ become less flat 
when 
the squeezing is increased, which can appear counterintuitive. It can be 
understood as a consequence of confinement-induced increase of interaction 
strength. Similar effect appears in very weakly bound helium dimers where 
significant enhancement of binding and decrease in size is predicted in the 
planar geometry when the width of the holding potential is approximately equal 
to the range of the potential~\cite{HeliumDimer}.  We have also shown that the
droplets with small number of particles significantly change their aspect ratio 
when the magnetic field is decreased ($|\delta a|$ increased) becoming 
significantly less flat for the weakest magnetic field (most strongly bound 
droplet) considered.
 
 As the number of particles in the cluster is increased, the central density
increases and reaches saturation, at which point the increase in confinement strength
makes the clusters extend perpendicularly to the direction of squeezing, again 
in a manner which is very similar to the behavior of Helium
clusters~\cite{he2d}. Bose-Bose saturated drops are observed only when the number of atoms
are orders of magnitude larger than $N_c$, hindering their experimental observation due
to three-body loses.
 
 Finally, our work points to the possibility of controlling droplet size and binding by
 adjusting the harmonic confinement in one direction. Increase of measurement 
precision would also enable discerning the effects beyond the LHY term, which are
most prominent for the $B=56.230$ G, corresponding to the most strongly bound
droplets, when they amount to around 25\% of the energy and central density.

\begin{acknowledgments}
We acknowledge financial support from Ministerio de Ciencia e Innovaci\'on 
MCIN/AEI/10.13039/501100011033
(Spain) under Grant No. PID2020-113565GB-C21 and from AGAUR-Generalitat de Catalunya Grant No. 2021-SGR-01411. The computational resources of the server UniST-Phy at the University of Split were used.
\end{acknowledgments}

%\bibliography{references.bib} % Path to your References.bib file

%apsrev4-2.bst 2019-01-14 (MD) hand-edited version of apsrev4-1.bst
%Control: key (0)
%Control: author (8) initials jnrlst
%Control: editor formatted (1) identically to author
%Control: production of article title (0) allowed
%Control: page (0) single
%Control: year (1) truncated
%Control: production of eprint (0) enabled
%
	
\end{document}